\documentclass[aps,twocolumn,superscriptaddress,showpacs,preprintnumbers,nofootinbib,amsmath,amssymb]{revtex4}
\usepackage{braket}
\usepackage{bm}
\usepackage{dcolumn}
\usepackage{graphicx}
\usepackage{color}
\usepackage[colorlinks=true,linkcolor=red,citecolor=blue]{hyperref}
\usepackage{aas_macros}
\usepackage{enumitem}
\usepackage{comment}

\begin{document}
\title{Dark age consistency  in the 21cm global signal}

\author{Fumiya Okamatsu}
\affiliation{Graduate School of Science and Engineering, Saga University, Saga 840-8502, Japan }

\author{Teppie Minoda}
\affiliation{Tsinghua University, Department of Astronomy, Beijing 100084, China}

\author{Tomo Takahashi}
\affiliation{Department of Physics, Saga University, Saga 840-8502, Japan}

\author{Daisuke Yamauchi}
\affiliation{Department of Physics, Faculty of Science, Okayama University of Science, 1-1 Ridaicho, Okayama, 700-0005, Japan}

\author{Shintaro Yoshiura}
\affiliation{Mizusawa VLBI Observatory, National Astronomical Observatory Japan, 2-21-1 Osawa, Mitaka, Tokyo 181-8588, Japan}

\date{\today}

\begin{abstract}
We propose a new observable for the 21cm global signal during the dark ages, {\em the dark-age consistency ratio}, which is motivated from the fact that the shape of the functional form of the brightness temperature against the frequency is cosmological-parameter independent in the standard $\Lambda$CDM model. The dark-age consistency ratio takes a certain definite value in the $\Lambda$CDM case, which can serve as a critical test of the model and probe those beyond the standard one. The new observable just needs measurements of the brightness temperature at a few frequency bands during the dark ages, and thus it allows us to test cosmological scenarios even with limited information on the global signal.

\end{abstract}



\maketitle

{\it Introduction---}
The 21cm line of neutral hydrogen can probe the evolution of the Universe over a wide range of redshift, especially at its higher end. Current observations of the global signal (sky-averaged signal) of the 21cm line such as EDGES \cite{Bowman:2018yin}, SARAS \cite{Singh:2021mxo} and its fluctuations such as LOFAR \cite{2019MNRAS.488.4271G,Gehlot:2020pul}, MWA \cite{Ewall-Wice:2016ysg,Yoshiura:2021yfx},  OVRO-LWA \cite{Eastwood:2019rwh} and so on have already reached high redshifts at the cosmic dawn. In the near future, SKAO \cite{SKAO} will also be operative on such high redshift with much better sensitivities.  Actually,  the 21cm signal can also appear even during the so-called dark ages ($ 30 \lesssim z \lesssim 150$), however, the observation of which is really challenging due to the Earth's ionosphere, the radio frequency interference (RFI), and so on. 

Indeed the dark ages have never been observed by any means, however this era lies prior to the formation of the first objects in the Universe,  and hence the pristine information of cosmology can be derived without  astrophysical uncertainties. The 21cm line of neutral hydrogen  would be a unique probe of this era and there have been theoretical works to study what aspects of cosmology can be probed by using its signal during the dark ages such as primordial power spectrum \cite{Loeb:2003ya,Balaji:2022zur}, primordial non-Gaussianities \cite{Cooray:2006km,Munoz:2015eqa,Floss:2022grj,Yamauchi:2022fri,Orlando:2023dgt}, isocurvature fluctuations \cite{Gordon:2009wx,Kawasaki:2011ze}, gravitational waves \cite{Book:2011dz,Hirata:2017dku}, primordial black holes \cite{Cole:2019zhu,Yang:2021agk,Yang:2022puh}, dark matter-baryon interaction \cite{Tashiro:2014tsa,Munoz:2015bca}, dark matter annihilation/decay \cite{Furlanetto:2006wp,Valdes:2007cu,Yang:2016cxm}, a test of statistical isotropy and homogeneity \cite{Shiraishi:2016omb}, neutrino masses \cite{Loeb:2003ya} and so on, in most of which fluctuations of the 21cm line are mainly studied although the global signal can also be useful.  

Currently, various observation plans are being discussed to probe the 21cm signal during the dark ages using a telescope on the moon or a satellite orbiting around the moon, which can avoid the Earth's ionosphere and the RFI and make its detection possible, such as FARSIDE \cite{FARSIDEweb,Burns:2021pkx},  DAPPER \cite{DAPPERweb,Burns:2021ndk}, NCLE \cite{NCLEweb,2020AAS...23610203C}, LCRT \cite{LCRTweb,Goel:2022jgw}, DSL \cite{Chen:2019xvd}, and so on (see also \cite{Koopmans:2019wbn}).  On the theoretical side, a precision cosmology using the dark age 21cm signal has been put forward by \cite{Mondal:2023xjx} (for an early work on the parameter estimation using the global signal during the dark ages, see \cite{Pritchard:2010pa}).  Although the 21cm fluctuations would bring a lot of information on various cosmological aspects, the first target of these lunar missions  would be the global signal. It is therefore a timely issue indeed to consider what we can learn from the 21cm global signal during the dark ages, having planned these missions in mind. 

\vspace{2mm}
{\it 21cm global signal---}
Here we give some basic formulas to calculate the 21cm signal, particularly during the dark ages.  For reviews of the 21cm physics, we refer the readers to e.g., \cite{Furlanetto:2006jb,Pritchard:2011xb}. The 21cm signal is characterized by the so-called differential brightness temperature $T_b$:
\begin{equation}
\label{eq:Tb}
 T_b  = \frac{T_s - T_\gamma}{1+z} \left(1 - e^{-\tau_\nu} \right) \,,
\end{equation}
where $T_s$ and $T_\gamma$ are the spin and radiation temperatures. Since we consider photons of cosmic microwave background (CMB) as a backlight, $T_\gamma$ coincides with the CMB temperature so that $T_\gamma = T_{\rm CMB}$. $\tau_\nu$ is the optical depth, which is given by 
\begin{equation}
\label{eq:tau}
\tau_\nu = \frac{3 c  h_p \lambda_{21}^2 A_{10} x_{\rm HI} (1 - Y_p) n_b }{32 \pi k_B T_s H} \,,
\end{equation} 
where $\lambda_{21}$ is the wave length for the 21cm line,  $A_{10}$ is the Einstein A coefficient for the spontaneous decay,  $x_{\rm HI}$ is the neutral fraction of hydrogen, $Y_p$ is the primordial mass fraction of helium, $n_b$ is the number density of baryon and $H$ is the Hubble parameter.  $c, h_p$ and $k_B$ are the speed of light, the Planck constant and the Boltzmann constant. By inserting Eq.~\eqref{eq:tau} to Eq.~\eqref{eq:Tb}, and assuming that the Universe is matter dominated and $\tau_\nu \ll 1$, one obtains 
\begin{eqnarray}
\label{eq:Tb2}
 && T_b  
  \simeq  
 85\,{\rm mK} ~ 
 \left( \frac{T_s -T_\gamma}{T_s} \right) 
 \left(  \frac{\omega_b}{0.02237} \right)  \notag \\
&& 
~~~~
\times \left( \frac{0.144}{\omega_m} \right)^{1/2} 
 \left( \frac{1-Y_p}{1-0.24} \right) 
 \left( \frac{1+z}{100} \right)^{1/2}  x_{\rm HI} \,.
 \end{eqnarray}
From this expression one can see that we just need to specify the following cosmological parameters to calculate $T_b$ in the standard $\Lambda$-Cold-Dark-Matter ($\Lambda$CDM) model):
baryon density $\omega_b\equiv \Omega_bh^2$, cold dark matter density  $\omega_m \equiv \Omega_ch^2$, with $\Omega_i$ the energy density of $i$-th component normalized by the critical energy density and $h$ the Hubble constant in units of $100\,{\rm km}/{\rm s}/{\rm Mpc}$.

\begin{figure}[htbp]
\begin{center}
     \includegraphics[width=8cm]{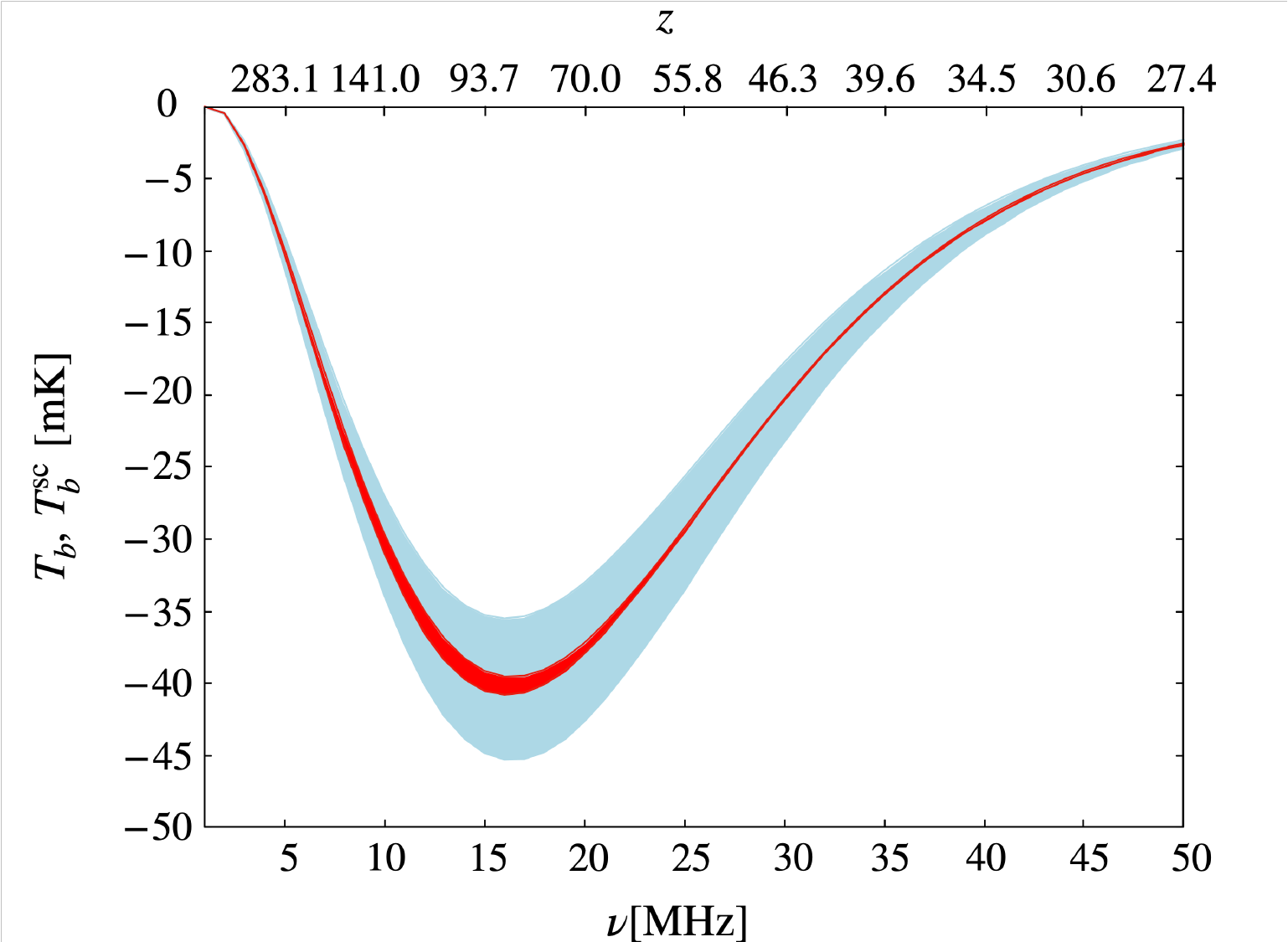}
\end{center}
\caption{$T_b$ with (without) the rescaling of Eq.~\eqref{eq:Tb_norm} is shown in red (blue). Here we vary $\omega_b$, $\omega_m$ and $Y_p$ within 5$\sigma$ ranges from Planck $\omega_b = 0.02237 \pm 0.00015,  \omega_c = 0.12 \pm 0.0012$ \cite{Planck:2018vyg} and Hsyu et al. $Y_p =0.2436 \pm 0.00395$ \cite{Hsyu:2020uqb}. \label{fig:Tb_rescaled}}
\end{figure}

The evolution of the spin temperature can be given by~\cite{1958PIRE...46..240F}
\begin{equation}
\label{eq:Ts}
T_s^{-1} = \frac{T_\gamma^{-1} + x_c T_k^{-1} + x_\alpha T_k^{-1}}{1 + x_c + x_\alpha} \,,
\end{equation}
where $T_k$ is the matter temperature and $x_c$ is the coefficient for atomic interactions, which is mainly determined by HH collisions during the dark ages and depends on the cosmological parameters as  $x_c \propto \omega_b (1-Y_p)$ since it depends on the number density of the  hydrogen. $x_\alpha$ is the coefficient for the Wouthuysen–Field effect \cite{1952AJ.....57R..31W,1958PIRE...46..240F}, in which Lyman-$\alpha$ photons effectively 
induce the transition between the hyperfine states. Actually it can be neglected during the dark ages in the standard $\Lambda$CDM case, although one needs to take it into account in some cosmological scenarios beyond the $\Lambda$CDM model.  
Notice that  Eq.~\eqref{eq:Ts} in the $\Lambda$CDM model during the dark ages gives, 
\begin{equation}
\label{eq:Tgamma_Ts}
\frac{T_s - T_\gamma}{T_s} 
= \frac{x_c }{1+x_c} 
\left( 1 - \frac{T_\gamma}{T_k} \right) \,,
\end{equation}
and in the later stage of the dark ages ($ 30 \lesssim z \lesssim 80$), $x_c \ll1 $ is realized, and then one can find the the scaling of the brightness temperature against the cosmological parameters as 
\begin{equation}
\label{eq:scaling}
T_b \propto \frac{\omega_b^2 (1-Y_p)^2}{\omega_m^{1/2}} \,.
\end{equation}
The scaling of $\omega_b$ and $\omega_m$ has also been noticed in \cite{Mondal:2023xjx}. 
Actually, by defining the following quantity  
\begin{equation}
C(\omega_b, \omega_m, Y_p) \equiv \frac{\omega_b^2 (1-Y_p)^2}{\omega_m^{1/2}} \,,
\end{equation}
and rescaling $T_b$ as 
\begin{equation}
\label{eq:Tb_norm}
T_b^{\rm sc} (\nu ;\widetilde{\bm\theta},{\bm\theta})
= T_b (\nu ; {\bm\theta}) 
\frac{C(\widetilde{\bm\theta})}{C({\bm\theta})} \,,
\end{equation}
where $\bm\theta = (\omega_b, \omega_m, Y_p)$, one can obtain an almost identical brightness temperature.
In Fig.~\ref{fig:Tb_rescaled}, we show the $T_b$ with and without the rescaling according to Eq.~\eqref{eq:Tb_norm}, where we varied the cosmological parameters in the range of 5$\sigma$ bounds from Planck data \cite{Planck:2018vyg} for $\omega_b$ and $\omega_m$, and that from Hsyu et al.~\cite{Hsyu:2020uqb} for $Y_p$~\cite{yp_comment}. For calculations of the brightness temperature, we used a modified version of {\tt recfast} \cite{Seager:1999bc,Seager:1999km,Wong:2007ym,Scott:2009sz}. As seen from the figure, the rescaled $T_b$ (red)  have almost the identical shape, due to the fact that $T_b$ scales as Eq.~\eqref{eq:scaling} and the position of the absorption trough remains unchanged even when the cosmological parameters are varied. On the other hand, $T_b$ without the rescaling (blue) are widely distributed.

\begin{figure}[htbp]
\begin{center}
     \includegraphics[width=8cm]{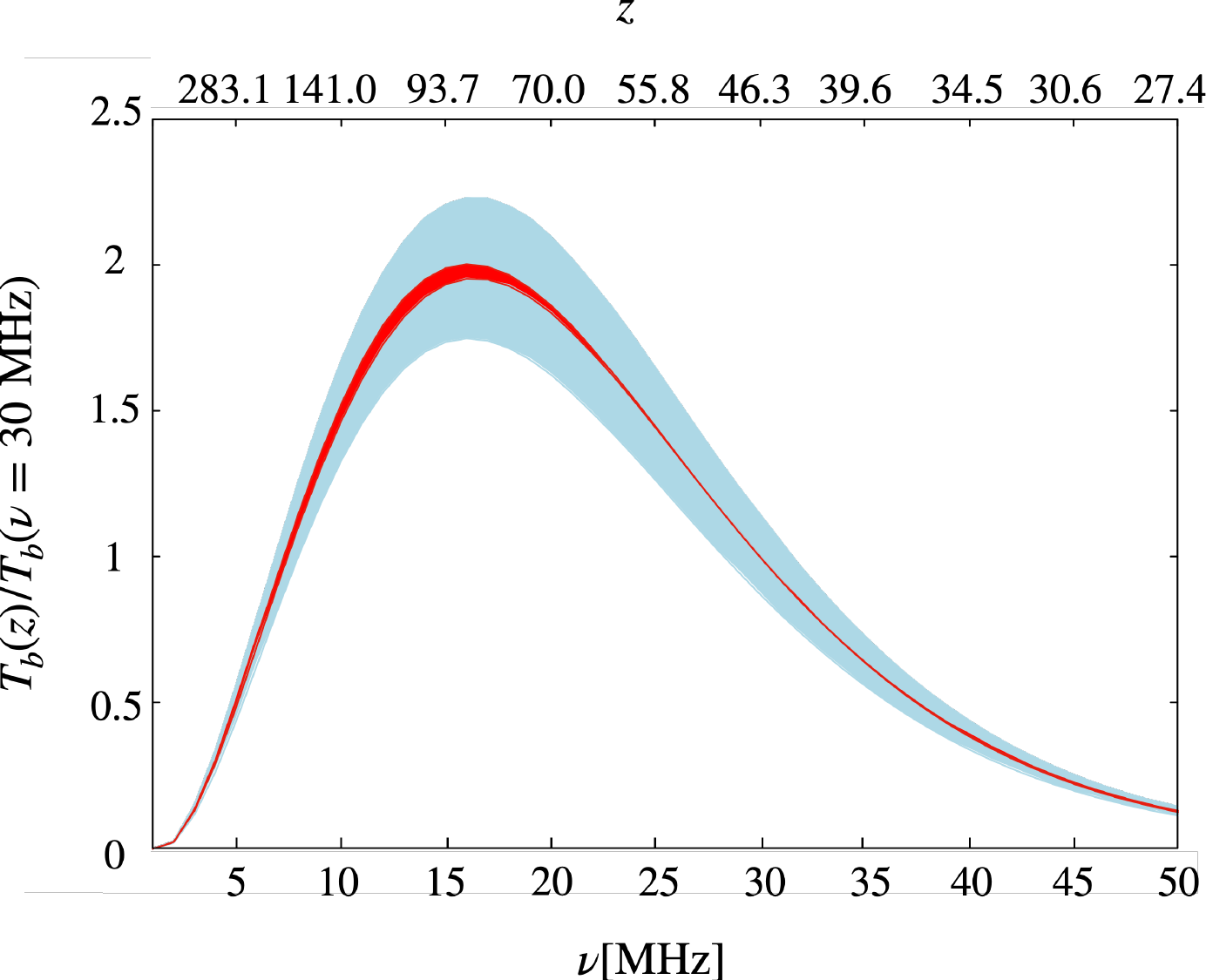}
\end{center}
\caption{Plots of $T_b(\nu) / T_b (\nu = 30\,{\rm MHz})$ with $\omega_b$, $\omega_m$ and $Y_p$ being varied within 5$\sigma$ ranges (red). For comparison, the case where $T_b(\nu)$ is just normalized by a constant as $T_b(\nu) /(-20.3\,{\rm mK})$ is also depicted (blue). 
 \label{fig:Tb_ratio}}
\end{figure}

To see more clearly that the shape of $T_b$ as a function of the frequency is cosmological-parameter independent in the $\Lambda$CDM model, in Fig.~\ref{fig:Tb_ratio}, we plot $T_b (\nu)$ divided by that at some reference frequency $\nu_\ast$ (red), which we take $\nu_\ast = 30\,{\rm MHz}$ for illustration purposes, with $\omega_b$, $\omega_m$ and $Y_p$ being varied within 5$\sigma$ ranges as done in Fig.~\ref{fig:Tb_rescaled}. For comparison, we also show $T_b(\nu)$ just divided by $-20.3~{\rm mK}$ (blue), which corresponds to the value of $T_b (\nu = 30\,{\rm MHz})$ for the case 
assuming the mean values for the cosmological parameters. As seen from the figure, different cosmological parameters give almost identical shapes for $T_b(\nu)$, particularly for the frequency range of $20\,{\rm MHz} < \nu < 50\,{\rm MHz}$, which corresponds to the later stage of the dark ages $ 30 \lesssim z \lesssim 80$.

\vspace{2mm}
{\it Consistency ratio as a new observable---}
The above arguments motivate us to consider the ratio of $T_b$ at two different frequencies, which should take an almost certain definite value  regardless of the cosmological parameters and can be used as a consistency check of the model. 

We define the ratio as 
\begin{eqnarray}
\label{eq:def_R}
R_{\nu_i / \nu_j} \equiv \frac{T_b(\nu=\nu_i\,{\rm [MHz]})}{T_b(\nu=\nu_j\,{\rm [MHz]})} \,,
\end{eqnarray}
where $\nu_i$ and $\nu_j$ are two different frequencies, which we call the ``{\em the dark-age consistency ratio}" since this ratio would remain to take the same value in the $\Lambda$CDM model to a high accuracy (although it is not exact) even when we vary the cosmological parameters, particularly for $\nu_i$ and $\nu_j$ being taken in the range of $20\,{\rm MHz} < \nu < 50\,{\rm MHz}$. In Table~\ref{tab:LCDM_ratio}, we show the ratios for several values of $\nu_i$ with the reference frequency $\nu_j = 30\,{\rm MHz}$.  As one can see from the table, the dark-age consistency ratios in the $\Lambda$CDM model are determined better than one percent accuracy regardless of the values of the cosmological parameters. Therefore if some observation indicates a deviation of $R_{\nu_i/\nu_j}$ from the prediction of the $\Lambda$CDM model, it suggests a model beyond the standard one. In particular, the consistency ratio proposed here would be very useful, even in the early stage of lunar missions mentioned in the introduction where the data of some limited frequency bands may be available. 
Even in such a case, the consistency ratio just needs measurements of $T_b$ at just two separate frequency bands. Detailed discussion on expected constraints on the consistency ratio in future missions will be given in a separate work \cite{Okamatsu_prep}.

\begin{table}
  \centering
  \begin{tabular}{|c|c|c|}
\hline
$\nu_i$ & $R_{\nu_i/30}$ & $T_{b}(\nu_{i})\,[{\rm mK}]$   \\ \hline 
$~~40~~$ &  $0.3873 \pm 0.0029$ ($0.76\%$)  & $-7.923\pm1.0107$ ($12.76\%$)   \\ 
\hline
$35$ & $0.6401 \pm 0.0016$ $(0.24\%)$ & $-13.10\pm1.7301$ $(13.20\%)$   \\
\hline
$25$ & $1.4454 \pm 0.0023$ $(0.16\%)$ & $-29.59\pm3.9133$  $(13.23\%)$ \\ 
\hline
$20$ & $1.8487 \pm 0.0126$ $(0.68\%)$  & $-37.82\pm4.8074$ $(12.71\%)$ \\
\hline
\end{tabular} 
  \caption{Ratio $R_{\nu_i/30}$ for several cases of $\nu_i$ in the $\Lambda$CDM model. The uncertainty refers to the variation when the cosmological parameters ($\Omega_bh^2, \Omega_mh^2, Y_p)$ are varied within the $5\sigma$ range. For comparison, the range of $T_b$ for the 5$\sigma$ variation of the cosmological parameters are also tabulated.
    \label{tab:LCDM_ratio}
  }
\end{table}

\vspace{2mm}
{\it Testing cosmology with the consistency ratio ---}
The consistency ratio defined in Eq.~\eqref{eq:def_R} should be useful to probe cosmological models since, 
as discussed above, it takes a definite constant value to a high accuracy during the dark ages in the $\Lambda$CDM model as shown in Table~\ref{tab:LCDM_ratio}. A possible deviation from the $\Lambda$CDM value can arise by violating (one or more) following assumptions during the dark ages: 
\begin{enumerate}[label=(\roman*),align=left]
    \item The Universe is matter-dominated.
    \item Lyman-$\alpha$ sources are negligible.
    \item Matter and photons are coupled via the Compton scattering.
    \item Radiation field is determined by CMB. 
\end{enumerate}

\begin{figure}[htbp]
\begin{center}
     \includegraphics[width=8cm]{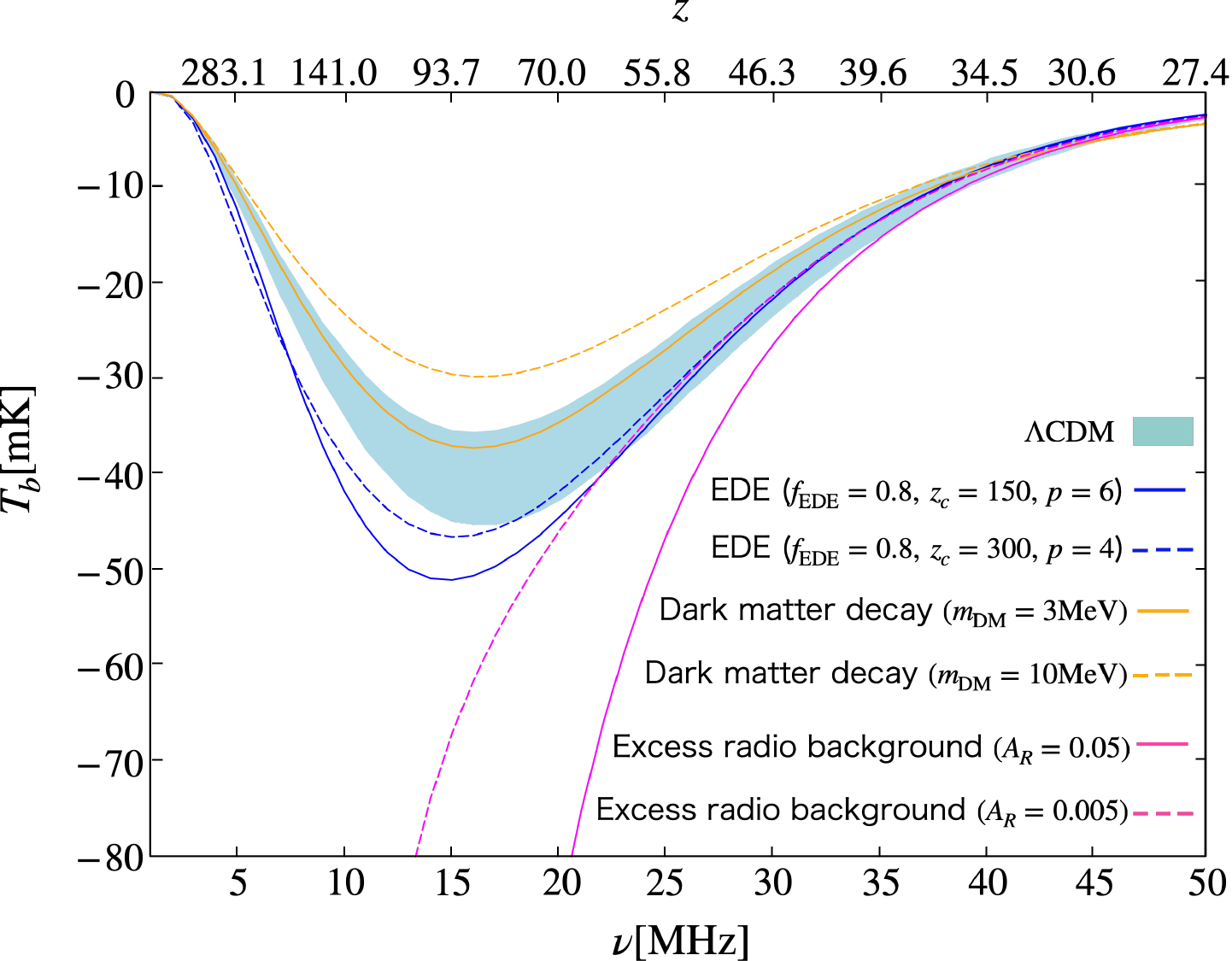}
\end{center}
\caption{Plots of $T_b$ for several models: EDE with $(f_{\rm EDE}, z_c, p) =(0.8, 150,6)$ and $(0.8, 300, 4)$ (blue solid and dashed), dark matter decay with $m_{\rm DM}= 3\,{\rm MeV}$ and $10\,{\rm MeV}$ (orange solid and dashed), excess radio background with $A_R=0.05$ and $0.005$  (purple solid and dashed).
The cosmological parameters are fixed as $\Omega_bh^2=0.02237,  \Omega_ch^2  = 0.12$ and $Y_p = 0.2436$. For reference, $T_b$ for the $\Lambda$CDM case is shown with the cosmological parameters varied within 5$\sigma$ ranges. 
 \label{fig:Tb_beyond} }
\end{figure}

An example of the violation of (i) is the so-called early dark energy (EDE) scenario where a component behaving like dark energy exists in some early times much before the current accelerating Universe. EDE has been attracting attention in several occasions such as a possible solution to the Hubble tension \cite{Kamionkowski:2022pkx,Poulin:2023lkg} (for the current status of the tension, see, e.g., \cite{DiValentino:2021izs,Perivolaropoulos:2021jda}). EDE may also be able to address the so-called Helium anomaly \cite{Takahashi:2022cpn} where the primordial Helium abundance measured by EMPRESS \cite{Matsumoto:2022tlr} may suggest a non-standard cosmological scenario.  Actually EDE has also been considered  to explain the EDGES signal \cite{Hill:2018lfx}, in which EDE can become a non-negligible component, or even a dominant one during the dark ages. 
To describe the energy density of EDE, we can consider the following functional form adopted in \cite{Hill:2018lfx}:
\begin{equation}
\rho_{\rm EDE} 
= C_{\rm EDE} \frac{1 + a_c^p}{a^p + a_c^p} \,,
\end{equation}
where $a_c$ is the scale factor at which the behavior of the EDE energy density changes from $\rho_{\rm EDE} = {\rm const.}$  to $\rho_{\rm EDE} \propto a^{-p}$. $C_{\rm EDE}$ can be fixed by giving the fraction of EDE at $a_c$ which is defined as 
\begin{equation}
\label{eq:fEDE}
f_{\rm EDE} 
= \left. \frac{\rho_{\rm EDE} (z)}{\rho_{\rm tot} (z)} \right|_{z=z_c} 
= \frac{\rho_{\rm EDE} (z_c)}{ \rho_{r, m, \Lambda} (z_c) + \rho_{\rm EDE} (z_c)}  \,,
\end{equation} 
where $ \rho_{r, m, \Lambda} (z)$ is the sum of energy densities of radiation, matter and the cosmological constant. In Fig.~\ref{fig:Tb_beyond}, we show $T_b$ in the EDE model with  $(f_{\rm EDE}, z_c, p) =(0.8, 150,6)$ and $(0.8, 300, 4)$ as examples.

The assumptions (ii) and/or (iii) can be violated, for instance, in models where dark matter (DM) annihilates or decays since DM annihilation/decay can produce photons in the energy range of Lyman-$\alpha$ and give an extra heating source for the gas temperature $T_k$. Indeed there have been many works regarding the effects of DM annihilation/decay on the 21cm signal, in particular see \cite{Furlanetto:2006wp,Valdes:2007cu,Yang:2016cxm} for its implications for the 21cm signal during the dark ages. In Fig.~\ref{fig:Tb_beyond}, we show $T_b$ in models with light DM decay for the mass of $3\,{\rm MeV}$ and $10\,{\rm MeV}$ for illustration which are calculated in the same manner as in \cite{Valdes:2007cu}. The details of the calculations and cases with some other scenarios will be given in a separate paper \cite{Okamatsu_prep}. Other examples of the violation of the assumption~(iii) include models with baryon-dark matter interaction \cite{Tashiro:2014tsa,Dvorkin:2013cea,Barkana:2018lgd}, primordial magnetic field \cite{Schleicher:2008hc,Sethi:2009dd,Minoda:2018gxj,Bera:2020jsg}, and so on, which have also been discussed in the context of the EDGES signal. 

The assumption~(iv) can be affected by extra radio 
background, and it has been discussed as a possible explanation of the EDGES signal \cite{Feng:2018rje,Yang:2018gjd,Pospelov:2018kdh,Fialkov:2019vnb,Moroi:2018vci}. Such an extra radio source is also suggested by ARCADE2 \cite{Fixsen:2009xn} and LWA1 \cite{Dowell:2018mdb}, whose results motivate the following parametrization for the radiation temperature \cite{Fialkov:2019vnb}:
\begin{eqnarray}
\label{eq:T_gamma_extra}
T_\gamma =  T_{\rm CMB,0} (1+z) \left[ 
1 + A_R  \left( \frac{\nu}{\nu_{\rm ref}} \right)^\beta
\right] \,.
\end{eqnarray}
Here $A_R$ is the relative size of the extra source to the CMB temperature at the reference frequency $\nu_{\rm ref}$ and $\beta$ describes the frequency dependence of the radiation.  We should note that the functional form should depend on the generation mechanism, and it may have some cut-off at some frequency. However, we adopt the form~\eqref{eq:T_gamma_extra} for illustration purpose. In \cite{Fialkov:2019vnb}, it has been suggested that the case with $A_R=5.7$, $\nu_{\rm ref}=78\,{\rm MHz}$ and $\beta=-2.6$ could explain the EDGES signal at $z=17$, which however would significantly distort $T_b$ at the dark ages.  In Fig.~\ref{fig:Tb_beyond}, $T_b$ for the cases with $A_R = 0.05$ and $0.005$ for $\nu_{\rm ref}=78\,{\rm MHz}$ and $\beta=-2.6$ are shown.

\begin{figure}[htbp]
\begin{center}
\vspace{2mm}
     \includegraphics[width=8cm]{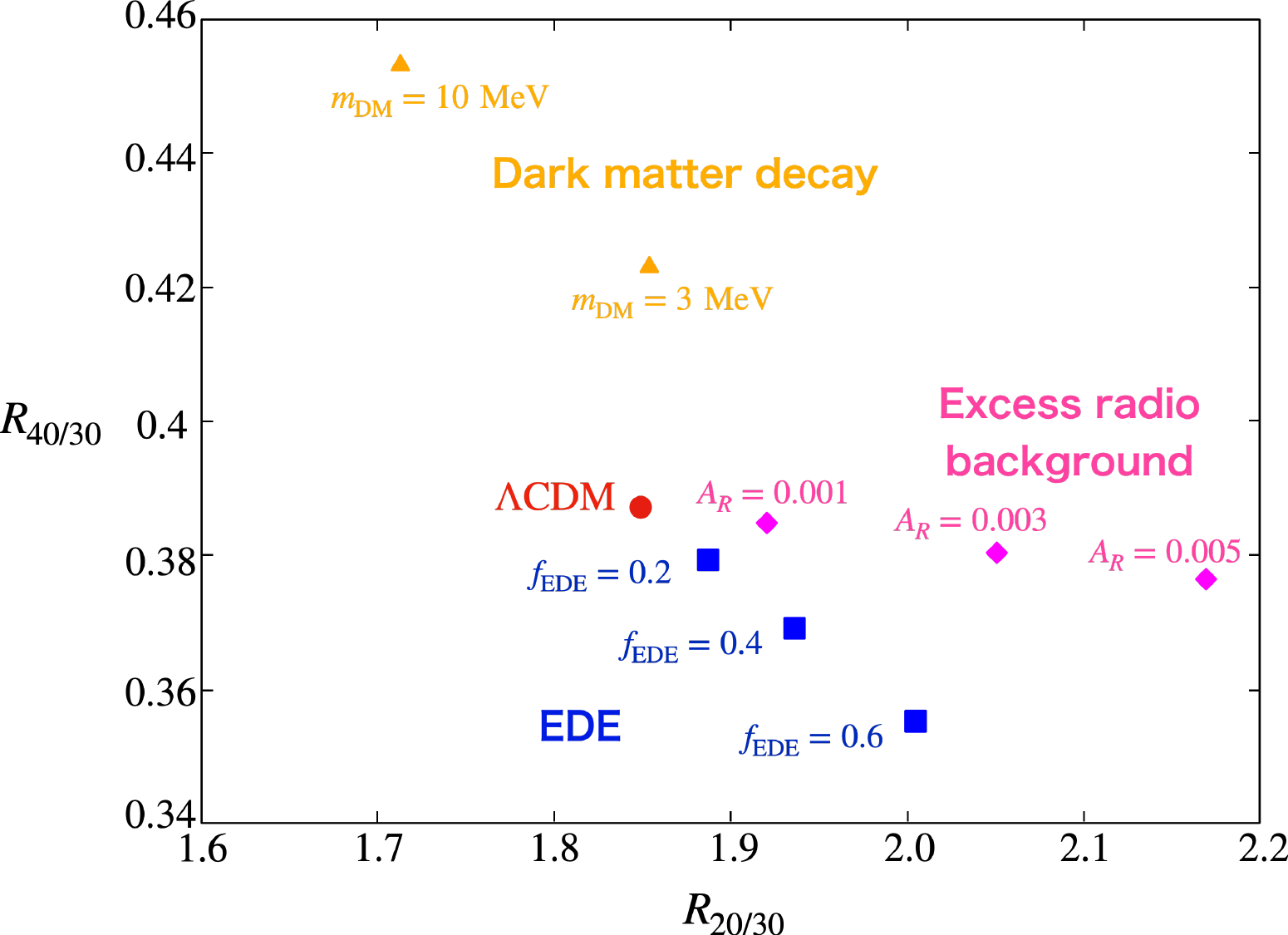}
\end{center}
\caption{Predictions for $R_{40/30}$ and $R_{20/30}$ of several example models are shown. Red point corresponds to the prediction of the $\Lambda$CDM model. Blue, orange and purple points are those for models with EDE ($p=4, z_c=150$), excess radio background ($\nu_{\rm ref} =78\,{\rm MHz}, \beta=-2.6$), and DM decay, respectively. Model parameters are depicted in the figure. 
 \label{fig:ratio}}
\end{figure}

Most models mentioned above can be tested by using the consistency ratio introduced in Eq.~\eqref{eq:def_R}.
In Fig.~\ref{fig:ratio}, the predictions of $R_{40/30}$ and $R_{20/30}$ for the $\Lambda$CDM model, and some other example models, such as EDE, 
excess radio background and DM decay, are shown. As discussed above, the $\Lambda$CDM model 
predicts certain definite values for the ratios with a very small uncertainty regardless of the values of the cosmological parameters, 
and its prediction is represented just by a point in the $R_{40/30}$--$R_{20/30}$ plane. In other models, 
the ratios are deviated from those of $\Lambda$CDM, from which one can clearly see that the new observable $R_{40/30}$--$R_{20/30}$ should be useful to probe models beyond the standard model. Notice that, when model parameters are varied, its predictions for $R_{\nu_i/\nu_j}$ also change.

\vspace{2mm}
{\it Conclusion ---}
We proposed a new observable for the 21cm global signal during the dark ages, {\em the dark-age consistency ratio}, which is motivated from the fact that the shape of $T_b$ as a function of the frequency is almost independent of the cosmological parameters in the $\Lambda$CDM model. Since it takes a certain definite value in the $\Lambda$CDM, it can be used as a consistency check of the model. If the deviation from the $\Lambda$CDM value is observed, it would signal a model beyond the standard scenario. The new observable only needs measurements at a few separate frequency bands, and hence fruitful information on cosmology can be derived from the dark age 21cm global signal even at the early stage of lunar or satellite missions in the foreseeable future.

\begin{acknowledgements}
This work was supported by JSPS KAKENHI 19K03874~(TT), 23K17691~(TT), 19H01891~(DY), 22K03627~(DY), 21J00416~(SY), 22KJ3092~(SY) and MEXT KAKENHI 23H04515~(TT). SY is supported by JSPS Research Fellowships for Young Scientists. This research was also supported by the grant of OML Project by the National Institutes of Natural Sciences (NINS program No, OML022303).
\end{acknowledgements}

\bibliography{refs}

\end{document}